\documentclass[12pt]{article} 		

\def\preprint{TUW-96/23}	
\def\finished{} 
\def\archive {hep-th/9701175}		

\def\title{	Calabi-Yau 4-folds and toric fibrations		}

\long\def\abstract{

	We present a general scheme for identifying fibrations in the 
	framework of toric geometry and provide a large list of weights for 
	Calabi--Yau 4-folds. We find 914,164 weights with degree
	$d\le150$ whose maximal Newton polyhedra are reflexive
	and 525,572 weights with degree $d\le4000$ that give rise to weighted
	projective spaces such that the polynomial defining a hypersurface
	of trivial canonical class is transversal.
	We compute all Hodge numbers, using Batyrev's formulas (derived
	by toric methods) for the first and Vafa's fomulas 
	(obtained by counting of Ramond ground states in $N=2$ LG models)
	for the latter class, checking their consistency for the 
	109,308 weights in the overlap.
	Fibrations of $k$-folds, including the elliptic case,
	manifest themselves in the $N$ lattice in the following simple way:
	The polyhedron corresponding to the fiber is a subpolyhedron
	of that corresponding to the $k$-fold, whereas the fan 
	determining the base is a linear projection of the fan 
	corresponding to the $k$-fold.

}

\let\Bbb=\mathbb 	\def\CY{Calabi--Yau}

\def\beq{\BE} \def\eeq{\EE} \def\eeql{\EEL} \def\bea{\BEA} \def\eea{\EEA}

\def\ip{\hbox{\bf 1}}\def\ipo{\hbox{\bf 0}} \def\IF{{\Bbb F}}\def\IW{{\Bbb W}} 

\def\fib{_{\rm fiber}} \def\bas{_{\rm base}} \def\Nt{{\tilde N}}

\usepackage{amssymb,latexsym,cite}		\catcode`\"=\active \let"=\"

\def\ifundefined#1{\expandafter\ifx\csname#1\endcsname\relax}

\def\HS#1 {\hspace*{#1pt}} \def\VS#1 {\vspace*{#1pt}} \long\def\del#1\enddel{} 
\def\BC{\begin{center}}	   \def\VR#1#2{\vrule height #1mm depth #2mm width 0pt}
\def\EC{\end{center}}	   \def\TVR#1#2{@{~~\VR{#1}{#2}}}	

\newfont{\XHbf}{cmbx10 scaled 4800}	\newfont{\XH}{cmr10 scaled 4800}

\let\0=\over	  \let\1=\vec	   \def\2{{1\over2}}	\let\3=\ss
\let\4=\underline \let\5=\overline \let\6=\partial	\def\7#1{{#1}\llap{/}}
\def\8#1{{\textstyle{#1}}}         \def\9#1{{\ifmmode{\pmb{#1}}\else\bf#1\fi}}

	  	   \def\({\left(}       \def\){\right)}

\def\EEL#1 {\label{#1}\EE}	\let\nn=\nonumber	
\def\BE {\begin{equation}} 	\def\EE {\end{equation}}	
\def\BEA{\begin{eqnarray}}	\def\EEA{\end{eqnarray}} 

\def\pmb#1{\setbox0=\hbox{${#1}$}   \kern-.025em\copy0\kern-\wd0
      \kern.05em\copy0\kern-\wd0     \kern-.025em\raise.0433em\box0 }
\let\Bbb=\mathbb 							
\def\IR{{\Bbb R}} \def\IC{{\Bbb C}} \def\IP{{\Bbb P}} 
 \def\IN{{\Bbb N}} \def\IZ{{\Bbb Z}}  

\def\mao#1{\mathop{\rm #1}\nolimits}	
\def\tr{\mao{tr}}

\let\and=\wedge		 		\let\ex=\times

\let\bra=\langle	\let\ket=\rangle	\def\<#1\>{\bra #1 \ket}

\def\rel#1 #2{\buildrel #1 \over {#2}}	

\def\BP{\begin{picture}} \def\EP{\end{picture}}		
\def\putlab#1)#2#3{\put#1){\makebox(0,0)[#2]{\small #3}}}
\def\putlin#1,#2,#3,#4,#5){\put#1,#2){\line(#3,#4){#5}}} 
\def\putvec#1,#2,#3,#4,#5){\put#1,#2){\vector(#3,#4){#5}}}
\def\putc#1)#2{\put#1){\makebox(0,0)[c]{#2}}}
\def\Bezier#1#2)#3)#4){\qbezier#2)#3)#4)} 	\thicklines

\def\subdef#1{\gdef\globalColor##1{##1}}	    
 	\subdef{Black}

			\let\d=\delta 	
 	 	\let\th=\theta 	
	\let\l=\lambda		
	 	\let\p=\pi 		\let\s=\sigma 
 	 	\let\c=\chi	 	
 	 	 	 	\let\S=\Sigma 
	 	 	 	\let\D=\Delta

 \def\cv{{\cal V}}

\def\plb#1 #2 {Phys. Lett. {\bf B#1} #2 }
\def\phr#1 #2 {Phys. Rep. {\bf  #1} #2 }	
\def\npb#1 #2 {Nucl. Phys. {\bf B#1} #2 }
\def\aph#1 #2 {Ann. Phys. {\bf #1} #2 }		
\def\jmp#1 #2 {J. Math. Phys. {\bf #1} #2 }
\def\jgp#1 #2 {J. Geom. Phys. {\bf #1} #2 }
\def\prd#1 #2 {Phys. Rev. {\bf D#1} #2 }
\def\prl#1 #2 {Phys. Rev. Lett. {\bf #1} #2 }
\def\rmp#1 #2 {Rev. Mod. Phys.  {\bf #1} #2 }
\def\zpc#1 {Z. Phys. {\bf #1C} }
\def\cmp#1 #2 {Commun. Math. Phys. {\bf #1} #2 }
\def\cqg#1 #2 {Class.Quant.Grav. {\bf #1} #2 }
\def\mpl#1 {Mod. Phys. Lett. {\bf A#1} }
\def\cpc#1 {Computer Phys. Commun. {\bf #1} }	
\def\ijmp#1 {Int. J. Mod. Phys. {\bf A#1} }
\def\ijmpC#1 {Int. J. Mod. Phys. {\bf C#1} }

\def\fnote#1#2{\begingroup\def\thefootnote{#1}\footnote{#2}
		\addtocounter{footnote}{-1}\endgroup}	

\def\offset#1#2{\def\xoff{#1}\def\yoff{#2}}     

\long\def\Fpar[#1](#2,#3)#4#5{\put(\xoff,\yoff){
                \put(#3,-#2){\makebox(0,0)[#1]{\parbox{#4mm}{#5}}}} }
\long\def\Fbox[#1](#2,#3)#4{\put(\xoff,\yoff){
                \put(#3,-#2){\makebox(0,0)[#1]{#4}}} }          \offset00

\newcounter{TRefNX} \let\OLDcite=\cite	\makeatletter
\def\makeTRefs#1{\@for 	\NewTRef:=#1\do{\global\makeTRef{\NewTRef}}}
\def\makeTRef#1{\ifundefined{TRef#1}\stepcounter{TRefNX}%
\expandafter\xdef\csname TRef#1\endcsname{\theTRefNX}\fi}\makeatother
\def\NEWcite#1{\makeTRefs{#1}\OLDcite{#1}}  

\ifundefined{draftmode} {} \else	\let\cite=\NEWcite
\newcount\HOUR 	\HOUR=\the\time \divide\HOUR by 60	\multiply \HOUR by 60 
\newcount\MIN	\MIN=\the\time 	\advance\MIN by -\HOUR 	\divide\HOUR by 60
\ifnum	\MIN>9	\def\printTIME{{\it\the\HOUR\,:\,\the\MIN}}
	\else	\def\printTIME{{\it\the\HOUR\,:\,0\the\MIN}} \fi 

\def\LLab#1{\BP(0,0)\unitlength=1mm\put(-12,.5){\makebox(0,0)[cr]{\small #1
        \rlap{$_{_{\makeatletter\csname TRef#1\endcsname\makeatother}}$}}}\EP}
\fi

\def\bye{\end{document}}		\long\def\new#1\endnew{{\bf #1}}
					\long\def\old#1\endold{{\bf #1}}
\begin{document}

\newfont{\XLbf}{cmbx10 scaled 2000}	\newfont{\XL}{cmr10 scaled 2000}
\newfont{\XLbfmath}{cmmi10 scaled 2000}

{\hfill \archive   \vskip -2pt \hfill\preprint \vskip -2pt \hfill UTTG-03-97}
\vskip 15mm
\centerline{\Huge\bf   \title }
\begin{center} \vskip 10mm
       	Maximilian KREUZER\fnote{*}{e-mail: kreuzer@tph.tuwien.ac.at}
\\[3mm]
	Institut f"ur Theoretische Physik, Technische Universit"at Wien\\
	Wiedner Hauptstra\3e 8--10, A-1040 Wien, AUSTRIA
\\[5mm]		              and
\\[5mm]	Harald SKARKE\fnote{\#}{e-mail: skarke@zerbina.ph.utexas.edu}
\\[3mm]	Theory Group, Department of Physics, University of Texas at Austin\\
	Austin, TX 78712, USA
	
\vfill                  {\bf ABSTRACT }	\end{center}	\abstract

\vfill \noindent \preprint\\[3pt] UTTG-03-97\\[5pt] \finished \vspace*{9mm}
\thispagestyle{empty} \newpage
\pagestyle{plain} 

\newpage

\section{Introduction}

It used to seem obvious from the critical dimension of superstrings and the
apparent dimension of space-time that only complex manifolds with dimensions
up to 3 play a role in string theory. 
The second string revolution, however, 
has changed this picture: It was shown
that strong coupling phenomena may increase the effective dimension of
space-time to 11 dimensions \cite{wit95}, 
thereby providing a geometrical interpretation 
of a number of string dualities in the context of M-theory.
Geometrization of the $SL(2,\IZ)$ symmetry of type IIB strings may even
lead to 12 dimensions \cite{F}.
Whether or not 
there are situations with an effectively 12-dimensional space-time, 
F-theory compactification already has an impressive record as a way to 
talk about compactifications of type IIB strings
and providing the missing 
geometrizations of duality symmetries \cite{F,F3,break,F4}. 
In a related development, non-perturbative physics of 3-dimensional 
compactifications seems to have the potential to teach us a lot about issues 
like SUSY breaking in 4 dimensions \cite{break},
which we may recover in some decompactification limit.

For all these issues it is important to have a number of generic 
examples of Calabi--Yau 4-folds at one's disposal
and, in particular for applications in F-theory, to know how to identify 
elliptic fibrations in an easy way.
The purpose of the present paper is twofold: 
On the one hand, we provide large classes of 4-folds in a systematic way.
In addition we give a detailed discussion of
the fibration structure in the toric context. 
{}From our experience with K3 fibrations \cite{k3} we expect that 
many families of toric \CY\ hypersurfaces will have members that
admit elliptic fibrations
in appropriate regions 
of the quantum moduli space (or, in technical terms, for
a triangulation of the fan that is compatible with the reflexive intersection
in the $N$ lattice that provides the fiber).

Toric methods are known to physicists mainly because of the work
of Batyrev \cite{ba93}, which appeared
in a situation where it had become 
inreasingly clear that complete intersections 
in products of (weighted) projective spaces were not general enough to grasp 
phenomena like mirror symmetry \cite{CICY,nms,kl94,aas}.
In the context of toric geometry, which  provides a natural extension of 
previous constructions, mirror symmetry 
manifests itself as
the elementary duality (or polarity)
of polytopes. 
The classification of toric Calabi-Yau hypersurfaces is
equivalent to the enumeration of reflexive 
polytopes, a problem that can be stated in simple combinatorial terms.

The link between the polytopes that generate the fan defining a toric variety
and weights that admit transversal polynomials of appropriate degree for 
Calabi--Yau hypersurfaces in weighted projective spaces turned out to be
very simple, at least in low dimensions: 
Just take the maximal Newton polytope (MNP), which consists of 
all exponent vectors of monomials whose degree is equal to the sum of weights. 
Its dual, which always contains a
simplex that encodes the weighted projective space we started with,
turns out to be integer for all MNPs in up to 4 dimensions, 
implying (by definition) that they are reflexive.
In the case of 4-dimensional hypersurfaces, however, we will see that 
only about 20\% of our MNPs are reflexive.

Even without transversality weight systems may lead to reflexive
polyhedra in the way we just described,
and any reflexive polyhedron is a subpolyhedron of an MNP
defined by one or several weight systems.
This observation is one of the keys to an approach for the classification 
of reflexive polyhedra \cite{crp,wtc}.
In the present context we used it to create large lists of weight systems
that lead to reflexive polyhedra. 
Here the fact that transversality and reflexivity are pracitcally 
unrelated properties in more than 4 dimensions becomes even more apparent:
Among the 914,164 weight systems we constructed that lead to reflexive
polyhedra, less than one percent allow for transversal polynomials!

Fibrations provide a beautiful example of how algebraic structures
in a toric variety manifest themselves in terms of linear structures
in the $N$ lattice.
As we will see, we can identify the fiber as a variety corresponding
to some subpolytope $\D^*\fib$ of $\D^*_{CY}$ and the base as a
variety whose toric description is given in terms of a fan $\S\fib$
that is a projection of $\S_{CY}$ along the direction of the sublattice
supporting $\D^*\fib$.
This makes it very easy to look for elliptic fibrations simply by 
looking for two dimensional integer subpolytopes containing the
interior point.
An even simpler approach would start from a multiply weighted space
such that one of the weight systems leads to the elliptic fiber.
Let us also mention here that our approach is particularly useful for 
discussing the degeneration of fibers.

In section 2 we discuss the relation between weighted projective spaces and
their toric resultions. 
We also present formulas and strategies for the
calculation of Hodge numbers.
In section 3 we discuss toric fibrations and the toric description of the
base manifold.
In section 4 we present 
our numerical results on weight systems leading to Calabi--Yau fourfolds.

While we were finishing the present work there appeared a preprint
\cite{KLRY} that partly overlaps with it.

\section{Weighted projective spaces vs. toric varieties }

We assume that the reader is familiar with basic notions of toric geometry,
such as the definitions of cones and fans 
(see, e.g., \cite{FU93} or \cite{OD88}).
We use standard notation, denoting the dual lattices by $M$ and $N$, their 
real extensions by $M_\IR$ and $N_\IR$, and the fan in $N_\IR$ by $\S$.
To each one-dimensional cone in $\S$ with primitive generator $v_k$ we 
assign a homogeneous coordinate \cite{cox} $z_k$, $k=1,\cdots,N$.
{}From the resulting $\IC^N$ we remove the exceptional set
\beq Z_\S=\bigcup_I\{(z_1,\cdots,z_N):\; z_i=0\;\forall i\in I\}\eeq
where the union $\bigcup_I$ is taken over all sets 
$I\subseteq \{1,\cdots,N\}$ for which 
$\{v_i:\;i\in I\}$ does not belong to a cone in $\S$.
Then our toric variety ${\cal V}_\S$ is given by the quotient of 
$\IC^N\setminus Z_\S$ by a group which is the product of a finite abelian 
group and $(\IC^*)^{N-n}$ acting by 
\beq (z_1,\cdots,z_N)\sim (\l^{w^1_j}z_1,\cdots,\l^{w^N_j}z_N)~~~~~
{\rm if}~~~~\sum_k w^k_j v_k=0 \eeql{er}
($N-n$ of these linear relations are independent).
Whenever $\S$ is simplicial,  the corresponding variety ${\cal V}_\S$ will 
have only quotient singularities.

Given a collection of positive integers $(w^1,\cdots, w^{n+1})$, 
called weights, there are two extreme ways of building a toric variety. 
In both cases one starts by taking $n+1$ vectors $v_i$ in $\IR^n$
such that any $n$ of them are linearly independent and $\sum w^iv_i=0$.
A convenient choice is 
\beq 
      v_i=e_i,~~~i=1,\cdots,n, ~~~~~ v_{n+1}=-  \8{1\0w^{n+1}}  \sum_1^n w^ie_i
\eeq
(or a multiple of these vectors).
The $M$ lattice is the lattice freely generated by the $v_i$.

The weighted projective space $\IW\IP_{(w^1,\cdots, w^{n+1})}$ is the 
toric variety determined by the unique fan whose one-dimensional cones
are $v_1,\cdots,v_{n+1}$.
The weights $w^i$ define a grading of monomials by 
$d(\prod z_i^{a_i})=\sum w^ia_i$. 
For the construction of Calabi--Yau hypersurfaces one considers 
quasi-homogeneous polynomials of degree $d=\sum w^i$.
A polynomial $W$ is said to be transverse if the set of equations $\6W/\6z_i=0$
is solved only by $z_i=0$ $\forall i$.
This condition ensures that the hypersurface has no singularities 
in addition to those coming from the singularities of the
ambient space.
These weight systems were classified for $n\le 4$ in \cite{nms,kl94}.

A sigma model on such a singular variety may be constructed as a 
particular phase of the low energy limit of an $N=2$ supersymmetric gauged 
linear sigma model \cite{wit93}.
A gauged linear sigma model with a single gauge field always 
contains a Landau--Ginzburg phase.
The numbers of chiral primary fields of a given charge in the $N=2$ 
superconformal field theory that is the low energy limit of a gauged 
linear sigma model do not change when going from one phase to another. 
Therefore we can use 
Vafa's formulas \cite{va89} for charge degeneracies in $N=2$ superconformal 
Landau--Ginzburg models to calculate what mathematicians call ``physicists' 
Hodge numbers''.

Another, a priori quite different approach, consists in
considering the (maximal) Newton polyhedron $\D$ associated with
the most general polynomial $W$ of degree $d=\sum w_i$.
This is just the convex hull of the points $(a_1,\cdots,a_{n+1})$ in 
$\IZ^{n+1}$ determined by the exponents occurring in the monomials of $W$.
By construction $\D$ lies in the hyperplane $\sum w^i a_i=d$ and contains 
the point $\ip=(1,\cdots,1)$.
After changing to $n$-dimensional integer coordinates, with $\ip\to \ipo$,
we may identify the resulting lattice with the $M$ lattice.
If the origin $\ipo$ of $M$ is in the interior of $\D$ (in this case
we say that the weight system has the ``interior point property''), the dual
polyhedron 
\beq \D^*=\{y\in N_\IR: \<y,x\>\ge -1 \forall x\in\D\}   \eeq
is bounded.
If, furthermore, all vertices of $\D^*$ are in $N$, $\D$ is said to 
be reflexive. 
The integer generators $v_1,\cdots v_{n+1}$ of above may be identified with
the points dual to the 	intersection of the 
planes $x_1=0,\cdots, x_{n+1}=0$ with the hyperplane given by the degree 
condition (before the change of coordinates).
$v_1,\cdots v_{n+1}$ are  points, but 
not necessarily vertices of $\D^*$.
If they are, then the coordinate hyperplanes correspond to
facets (codim 1 faces) of $\D$, i.e. the points of $\D$ affinely
span these hyperplanes.
In this case we say that a weight system has the ``span property''.
\del	In any case, reflexivity of $\D$ implies that the points $v_i$ belong
	to $\D^*\cap N$.
\enddel

It was shown in \cite{wtc} that transversality always implies the
interior point property 
and that for 
$n\le 4$ the interior point property implies reflexivity. 
The fact that transversality implies reflexivity of $\D$ for $n\le 4$ had
been checked by computer \cite{ca95,klun}. 
Note, however, that the proof of \cite{wtc} also applies to the much
larger class of all abelian
orbifolds \cite{aas} and the MNPs on the respective sublattices that 
arise by dividing out phase symmetries that still admit transversal 
polynomials.

Denoting the various sets of weight systems (in obvious notation)
by $T$, $I$ and $R$, the following relations between the different types 
hold:\\[4pt]
$n=2,3:\;T=I=R$,\\
$n=4:\:T\subset I=R$,\\
$n>4:\;T\subset I,\;R\subset I$, no further relations.\\[4pt]
The statement $T=I$ for $n=2,3$ is only known due to explicit
constructions of the corresponding sets of 3 or 95 weights, respectively
\cite{reid,wtc}.
In more than 2 dimensions the span property is independent of whether 
the weight system belongs to $T$, $I$ or $R$.

For a reflexive polyhedron $\D$ we now consider the
fan $\S$ over some triangulation of the faces of $\D^*$.
A Calabi--Yau hypersurface in $\cv_\S$ is given by the zero locus of
\beq p=\sum_{x\in \D\cap M}a_x\prod_{k=1}^N z_k^{\<v_k,x\>+1}.  \eeql{polyn}
If $\S$ is defined by a maximal triangulation of $\D^*$, the generic
hypersurface of this type is smooth for $n\le 4$ \cite{ba93}.
Also by \cite{ba93}, the Hodge numbers $h_{11}$ and $h_{1,n-2}$ are known,
and in \cite{ba94} the remaining Hodge numbers of the type $h_{1i}$
were calculated. 
For a hypersurface of dimension $n-1\ge 3$ these formulas can be summarised as
\bea h_{1i}&=&\d_{1i}\(l(\D^*)-n-1-\sum_{{\rm codim }\th^* =1}l^*(\th^*)\)+
	\d_{n-2,i}\(l(\D)-n-1-\sum_{{\rm codim }\th =1}l^*(\th)\)\nn\\
           &&+ \sum_{{\rm codim }\th^* =i+1}l^*(\th^*)l^*(\th) \label{baho}
                                                                       \eea
for $1\le i\le n-2$, where $l$ denotes the number of integer points of a 
polyhedron and $l^*$ denotes the number of interior integer points of a face.

For Calabi--Yau
4-folds there is a linear relation among the Hodge numbers that has been
obtained using index theorems in \cite{se96,KLRY}.
The same relation can, in fact, be obtained as a simple consequence of a 
sum rule for
charge degeneracies of Ramond ground states that has been derived from modular
invariance of the elliptic genus for arbitrary $N=2$ superconformal field 
theories \cite{ah94}:
\BE
	\tr (-)^F J_0^2=\8{c\036} \tr (-)^F=-\8{d\012}\c,
\EE
where $J_0$ is the (left-moving) U(1) charge, $c=3d$ is the central 
charge and the trace extends over the Ramond ground states
(we need to be careful with the sign of the Euler characteristic because the
Hodge numbers $h_{pq}$ of the $\s$ model on a Calabi-Yau manifold and the 
charge degeneracies $n_{pq}$ of Ramond ground states of charge 
$(Q_L,Q_R)=(p-{d\02},q-{d\02})$ are related by $h_{p,q}=n_{d-p,q}$).
For a Calabi--Yau manifold of arbitrary dimension this implies
\BE
	\sum_{p,q=0}^d(-)^{p+q}(p-\8{d\02})^2h_{p,q}=-\8{d\012}\c.
\EE
In 4 dimensions this equation is equivalent to 
\BE
	h_{22}=44+4h_{11}-2h_{12}+4h_{13}, 
\EEL{h22}
where we used Poincar\'e and Hodge duality of the Hodge diamond and omitted
the contribution $20h_{02}-52h_{01}$ on the r.h.s. 
	that vanishes for toric Calabi-Yau hypersurfaces. 
	For the Euler characterictic of 4-folds we thus find
\BE
	\c=6(8+h_{1,1}-h_{1,2}+h_{1,3})
\EE
In 2 dimensions the above sum rule uniquely determines the Hodge diamond of 
the K3 surface, whereas it is trivially satisfied (and therefore no so 
well-known) for CY 3-folds.

As an example for different cases that can occur in different dimensions
consider the weight system $(1,\cdots,1,2)$.
The weighted projective space $\IW\IP_{(1,\cdots,1,2)}$ can
be represented by the vectors 
\beq v_1=(1,0,\cdots,0),\;\cdots,\;v_n=(0,\cdots,0,1),\;
     v_{n+1}=(-1/2,\cdots,-1/2)   \eeq
in the lattice $N$ consisting of points in $\IR^n$ with coordinates
that are either all integer or all  half integer.
Independently of $n$, $\IW\IP_{(1,\cdots,1,2)}$ has precisely one pointlike
singularity, located
at $z_1=\cdots =z_n=0$ and determined by the $\IZ_2$ quotient 
$(z_1,\cdots,z_{n+1})\sim(-z_1,\cdots,-z_n,z_{n+1})$.
In terms of toric geometry, this singularity corresponds to
the fact that the simplex spanned by $0,v_1,\cdots,v_n$ has 
twice the volume of the unit simplex.
This singularity can always be resolved by blowing up the singular
point, i.e. by introducing an exceptional divisor (for a nice description
of the blowup of orbifold singularities in a recent review for physicists, 
see \cite{as96}).
In terms of toric geometry, this means that we subdivide the
cone generated by $v_1,\cdots ,v_n$ by adding an extra generator
\beq v_{n+2}:=(1/2,\cdots,1/2)=-v_{n+1}.   \eeq
As we will see, however, the fan over $\D^*$ does not necessarily
correspond to this type of desingularisation.

The $M$ lattice can be identified with the integer points in $\IZ^n$
such that $\sum x_i=0$ mod 2.
For even $n$, $W$ can be chosen as the Fermat polynomial 
\beq
	 z_1^{n+2}+\cdots+z_n^{n+2}+z_{n+1}^{(n+2)/2}, 
\eeq
so that the maximal Newton polytope $\D$ is a simplex,
whereas for odd $n$ the vertex corresponding to $z_{n+1}^{(n+2)/2}$
is replaced by $n$ vertices corresponding to expressions 
$z_{n+1}^{(n+1)/2}z_i$ with $i=1,\cdots,n$.
Let us now consider what happens for various values of $n$:\\[4pt]
$n=2$: The vertices of $\D^*$ are just $v_1$, $v_2$ and $v_3$. 
$v_4$ is in the interior of the edge (facet) $\overline{v_1v_2}$.
Whether we blow up $\IW\IP_{(1,1,2)}$ by the divisor corresponding
to $v_4$ does not matter because this divisor does not intersect
the hypersurface.\\[4pt]
$n=3$: The monomials $z_4^2z_i$ correspond to a plane in the $M$ lattice 
whose dual is $v_5$. 
Thus ${\cal V}_\S$ is the blow-up of $\IW\IP_{(1,1,1,2)}$.\\[4pt]
$n=4$: Once again $W$ is of Fermat type, but now $v_6$ lies outside $\D^*$.
The variety ${\cal V}_\S=\IW\IP_{(1,1,1,1,2)}$ has the $\IZ_2$ singularity,
but the generic hypersurface of degree 6 does not intersect it.
The blow-up of $\IW\IP_{(1,1,1,1,2)}$ corresponds to a different reflexive
polyhedron leading to a hypersurface with different Hodge numbers.\\[4pt]
$n=5$: The monomials $z_6^3z_i$ correspond to a plane in the $M$ lattice
whose dual is the point $v_7/2 \in N_\IR$ which is not in $N$.
$\D^*$ is no longer reflexive.
As there is no Fermat type monomial in $z_6$, any degree 7 hypersurface
in $\IW\IP_{(1,1,1,1,1,2)}$ intersects the singular point $z_1=\cdots=z_6=0$.
Vafa's formulas give $h_{11}=1$, $h_{12}=0$ and $h_{13}=455$.
They certainly do not correspond to the blow-up of $\IW\IP_{(1,1,1,1,1,2)}$:
Whereas the convex hull $\tilde\Delta^*$ of $v_1,\cdots,v_7$ is reflexive 
and the corresponding variety even smooth, $l(\tilde\Delta^*)$ being 8,
eq. (\ref{baho}) tells us that $\tilde h_{11}=8-6=2$.
We note, however, that eq. (\ref{baho}) can be applied to non-integer 
polyhedra as well.
In the present case the results of inserting $\Delta, \D^*$ into this
formula coincide with those of Vafa's formulas.
We do not know whether this is always true.

\section{Fibrations}

The aim of this section is to give a general recipe for identifying
fibrations of hypersurfaces of holonomy $SU(n-1)$ in $n$-dimensional
toric varieties where the generic fiber is an $n'-1$ dimensional
variety of holonomy $SU(n'-1)$.
In other words, it will apply to elliptic fibrations of
K3 surfaces, CY threefolds, CY fourfolds, etc., to K3 fibrations of
CY $k$-folds with $k\ge 3$, to threefold fibrations of fourfolds, and so on.
The main message is that the structures occurring in the fibration
are reflected in structures in the $N$ lattice: 
The fiber, being an algebraic subvariety of the whole space,
is encoded by a polyhedron $\D^*\fib$ which is a subpolyhedron of
$\D^*_{CY}$, whereas the base, which is a projection of the 
fibration along the fiber, can be seen by projecting the $N$ lattice
along the linear space spanned by $\D^*\fib$.
The details given in the following are somewhat technical; the reader
is advised to check the various steps with some explicit example,
e.g. the one given later.

Assume that $\D^*$ contains a lower-dimensional reflexive 
subpolyhedron $\D^*\fib=(N\fib)_\IR\cap \D^*_{CY}$ with the same interior
point.
This allows us to define a dual pair of exact sequences
\beq 0\to N\fib \to N_{CY}\to N\bas \to 0\eeql{esN}
and
\beq 0 \to M\bas \to M_{CY} \to M\fib \to 0.\eeql{esM}
Using the same arguments as in \cite{k3}, we can convince ourselves
that the image of $\D_{CY}$ under $M_{CY} \to M\fib$ is dual to $\D^*\fib$.
Let us also assume that the image $\S\bas$ of
$\S_{CY}$ under $\p: N_{CY}\to N\bas$ defines a fan in $N\bas$.
This is certainly not true for 
 arbitrary
triangulations of $\D^*$. 
Constructing fibrations, one should rather build a fan $\S\bas$
from the images of the one-dimensional cones in $\S_{CY}$
and try to construct a triangulation of $\S_{CY}$ and thereby of
$\D^*_{CY}$ that is compatible with the projection.
It would be interesting to know whether this is always possible
whenever the intersction of a reflexive polyhedron with a linear 
subspace of $N_\IR$ is again reflexive.

The set of one-dimensional cones in $\S\bas$ is the set of 
images of one-dimensional cones in $\S_{CY}$ that do not lie in $N\fib$. 
The image of a primitive generator $v_i$
of a cone in $\S_{CY}$  is the origin or
a positive integer multiple of a primitive
generator $\tilde v_j$ of a one-dimensional cone in $\S\bas$. 
Thus we can define a matrix $r^i_j$, most of whose elements are $0$, through
$\p v_i=r_i^j\tilde v_j$ with $r_i^j\in \IN$ if $\p v_i$ lies in
the one-dimensional cone defined by $\tilde v_j$ and $r_i^j=0$ otherwise.
Our base space is the multiply weighted space determined by
\beq (\tilde z_1,\cdots,\tilde z_\Nt)\sim (\l^{\tilde w^1_j}
\tilde z_1,\cdots,\l^{\tilde w^\Nt_j}\tilde z_\Nt),~~~~~~
j=1,\cdots,\tilde N-\tilde n  \eeql{ber} 
where the $\tilde w^i_j$ are any integers such that 
$\sum_i\tilde w^i_j \tilde v_i=0$.
The projection map from $\cv_\S$ (and, as we will see, from the Calabi--Yau
hypersurface) to the base is given by
\beq \tilde z_i=\prod_jz_j^{r_j^i}. \eeq
This is well defined: $z_j\to \l^{w_k^j}z_j$ leads to 
$\tilde z_i\to \l^{w_k^j r_j^i}\tilde z_i$ which is among the good
equivalence relations because applying $\p$ to $\sum w^j_k v_j=0$
gives $\sum w^j_k r_j^i \tilde v_i=0$.

A generic point in the base space will have $\tilde z_i\ne 0$ for all $i$,
implying $z_i\ne 0$ for all $v_i\not\in \D^*\fib$.
The choice of a specific point in $\cv_{\S\bas}$ and the use of all 
equivalence relations except for those involving only $v_i\in \D^*\fib$ 
allows to fix all $z_i$ except for those corresponding to 
$v_i\in \D^*\fib$.
Thus the preimage of a generic point in $\cv_{\S\bas}$
is indeed a variety in the moduli space determined by $\D^*\fib$.

What we have seen so far is just that $\cv_\S$ is a fibration over
$\cv_{\S\bas}$ with generic fiber $\cv_{\S\fib}$ (this is actually
the statement of an exercise on p. 41 of ref. \cite{FU93}) and
how this fibration structure manifests itself in terms of homogeneous
coordinates.
Now we also want to see how this can be extended to hypersurfaces.
To this end note that if $v_k\in\D^*\fib$ then $\<v_k,x\>$ only
depends on the equivalence class $[x]\in M\fib$ of $x$ under 
\beq x\sim y ~~~\hbox{ if }~~~ x-y \in M\bas . \eeq
Thus we may rewrite eq. (\ref{polyn}) as
\beq p=\sum_{[x]\in \D\fib\cap M\fib}a'_{[x]}
      \prod_{v_k\in\D^*\fib} z_k^{\<v_k,[x]\>+1}~~~\hbox{ with }~~~
    a'_{[x]}=\sum_{x\in [x]}a_x\prod_{v_k\not\in\D^*\fib}z_k^{\<v_k,x\>+1}.  
                                                                \eeql{polynf}
In each coordinate patch for $\cv_{\S\bas}$ this is just an equation
for the fiber with coefficients that are polynomial functions of
coordinates of the base space.

There are two different occasions upon which the fiber may degenerate:
The fiber being a hypersurface in $\cv_{\S\fib}$, it can either happen
that $\cv_{\S\fib}$ itself degenerates or that the coefficients
of the equation determining the fiber hit a singular point in the moduli space.
While we do not know any way to read off the occurrence of the second case
from the toric data, we can surely see the first case: 
Whenever a one-dimensional cone (with primitive generator $\tilde v_i$) 
in $\S\bas$ is the image of more than one one-dimensional cone in $\S$, 
the fiber becomes reducible over the divisor $\tilde z_i=0$ determined
by $v_i$.
Different components of the fiber correspond to different equations
$z_j=0$ with $\p v_j=r^i_j \tilde v_i$.
The intersection patterns of the different components of the
reducible fibers are crucial for understanding enhanced gauge symmetries
\cite{wit95,as96} and deserve further study.

Let us consider as an example the well known class of Calabi--Yau
hypersurfaces (threefolds) of degree $6n+12$ in $\IW\IP_{(1,1,n,2n+4,3n+6)}$
\cite{mvI}.
If $12/n$ is integer, the corresponding Newton polyhedron
will be a simplex and the dual polyhedron $\D^*$ can be described
as the convex hull of the points 
\beq (1,0,0,0),\;(0,1,0,0),\;(0,0,1,0),\;(0,0,0,1),\;(-1,-n,-2n-4,-3n-6)\eeq
in $N\simeq \IZ^4$.
The elliptic fiber is determined by $\D\fib^*$ with vertices
\beq (0,0,1,0),\;(0,0,0,1),\;(0,0,-2,-3)\;
\hbox{ in }\; N\fib=N\cap\{x_1=x_2=0\}. \eeq
Of course it is just the torus given by the Weierstrass equation
in $\IW\IP_{(1,2,3)}$.
The projection of
$N$ to $N\bas=N/N\fib$ is realised as $\p: (x_1,x_2,x_3,x_4)\to(x_1,x_2)$
(``throwing away the last two coordinates of each point'').
$\D^*$ gets projected to the convex hull of $(1,0)$, $(0,1)$ and $(-1,-n)$. 
Each of these points provides a one-dimensional cone in $\S\bas$,
which clearly is the fan of the Hirzebruch surface $\IF^n$ (compare,
for example, with \cite{FU93}, p. 7).
All other integer points are of the form $(0,-l)$ with integer $0\le l\le n/2$.
They correspond to the one-dimensional cone generated by $(0,-1)$ and give
examples for nontrivial values of the $r_i^j$ defined above.

Let us get even more specific and consider the case $n=4$.
Neglecting 4 points in the interiors of facets of $\D^*$ (the corresponding
divisors do not intersect the Calabi--Yau hypersurface and are therefore 
irrelevant for the present discussion), we can arrange
the integer points of $\D^*$ according to their images 
\beq \tilde v_1 := (1,0),~~ \tilde v_2 := (0,1), ~~\tilde v_3 := (-1,-4),~~
     \tilde v_4 := (0,-1),~~ \ipo:=(0,0)     \eeq
in $\S\bas$:
\bea 
\p v =& r \tilde v_1  &\hbox{ for } v_1 :=(1,0,0,0),\nn\\
\p v =& r \tilde v_2  &\hbox{ for } v_2 :=(0,1,0,0),\nn\\
\p v =& r \tilde v_3  &\hbox{ for } v_3 :=(-1,-4,-12,-18),\nn\\
\p v =& r \tilde v_4  &\hbox{ for } v_4 :=(0,-1,-3,-4),~~
                       v_5:=(0,-1,-4,-6),~~  v_6:=(0,-2,-6,-9),\nn\\
\p v =& \ipo          &\hbox{ for } v_7 :=(0,0,1,0), ~~v_8:=(0,0,0,1),~~
                                     v_9 :=(0,0,-2,-3).   \eea
The projection to the base is given by 
\beq
\tilde z_1=z_1,~~\tilde z_2=z_2,~~\tilde z_3=z_3,~~\tilde z_4=z_4z_5z_6^2.
\eeq
There are many linear relations among the $v_i$. 
One of them is $2v_7+3v_8+v_9=0$, ensuring that
\beq (z_1,\cdots,z_6,z_7,z_8,z_9) \sim (z_1,\cdots,z_6,\l^2z_7,\l^3z_8,\l z_9).
\eeq
With respect to this relation, $p$ is quasi-homogeneous of degree 6.
With linear redefinitions of $z_7,z_8,z_9$, one can bring $p$ into
Weierstrass form 
\beq p=z_8^2-z_7^3+f(z_1,\cdots,z_6)z_7z_9^4+g(z_1,\cdots,z_6)z_9^6.\eeq

Let us also briefly discuss strategies for finding fibrations.
One strategy is to take reflexive polyhedra and look for lower-dimensional
reflexive polyhedra that are contained in them.
In particular, looking for elliptic fibrations, one might just intersect
$\D^*$ with any two-dimensional plane spanned by integer points of $\D^*$.
Checking $N\choose 2$ pairs of points w.r.t. whether the plane spanned by them
carries a reflexive subpolyhedron is no challenge to present day 
computer power, even for large numbers of polyhedra.

An even simpler approach for constructing large numbers of
fibrations could make use of the following observation:
If $\D',\D''$ are reflexive polyhedra in lattices $M',M''$ respectively,
then 
\beq \D:=\D'\times\D''=\{(x',x''):~x'\in \D',~~x''\in \D''\}   \eeq
is also a reflexive polyhedron. Its dual is
\beq \D^*=\{(\l y',(1-\l)y''):~~y'\in \D',~~y''\in \D'',~~0\le\l\le 1\}   \eeq
and the set $\S_{(k)}$ of $k$-dimensional cones in $\S$ is given by
\beq \S_{(k)}=\{(v',v''):~~v'\in\S_{(k')}',~~v''\in\S_{(k'')}'',~~k'+k''=k\}.
     \eeq
Of course $\cv_\S=\cv_{\S'}\times\cv_{\S''}$, so $\cv_\S$ is trivially a 
fibration.
The equation $p=0$ defines a Calabi-Yau ($k-1$)--fold which may be
interpreted either as a fibration with base $\cv_{\S'}$ and generic
fiber a hypersurface in $\cv_{\S''}$ or as a fibration with base $\cv_{\S''}$ 
and generic fiber a hypersurface in $\cv_{\S'}$.
In this way one can immediately construct almost 3 million elliptic
fibration fourfolds by combining one of the 16 two-dimensional reflexive
polyhedra with reflexive polyhedra coming from one of the 
184,026 weight systems of \cite{wtc}.

\section{Scans for fourfolds}

We performed systematic searches for the two different types of 
Calabi--Yau fourfold models described by a single weight system
as analysed in section 2.
In both cases we used improved versions of the same basic strategy,
namely to check for all partitions $w^1,\cdots,w^{n+1}$ of
$d=6,7,\cdots$ whether they allow for transverse polynomials or
reflexive polyhedra. 

For the transverse weights, the preselection that we imposed to reduce the
number of partitions is, in the language of \cite{cqf}, ``compatibility with 
at most one unresolved pointer to an unknown weight''.
With this method
we obtained the complete list of 525572 weight systems of degree up to 4000. 
Based on the statistics given in table 1 we cannot give an estimate of the 
total number, since  $2\ex10^5/d$ is a good approximation to the average
number of transversal weights per degree, which would lead 
to a divergent sum. It is well-known, however, that this set is finite
\cite{cqf}.
A complete enumeration is impossible with our
present approach (this can be inferred from the rate at which our program slows
down with growing $d$ and the fact that the degrees of the 
3462 Fermat weights range up to $d=3263442$), but the numbers in
table 1 do not seem to exclude the possibility of a complete 
classification along the lines of refs. \cite{nms,kl94}.

The situation is quite different for 4-folds 
coming from reflexive weights, which are far more numerous. 
Here we are limited by disk space rather than by calculation time.
We have the complete list of 914164 weights with degree $d\le150$, only 
6918 of which are transversal.
Together with the weights that are both transversal and reflexive we 
have thus accumulated more than $10^6$ reflexive weights. 
Various sublists, as well as the files with the complete results are 
available on the internet.%
\footnote{
	The URL is http://tph.tuwien.ac.at/{\tiny$^\thicksim$}kreuzer/CY
}

A central goal of our computer studies was to investigate the relation 
between transversality and reflexivity of the maximal Newton polyhedra.
Whereas transversality always implies reflexivity in up to 4 
dimensions, it turned out that only about 20\% of the five dimensional
polyhedra defined by our transversal weights are reflexive.
This number shows only little dependence (in the form of a slight 
decrease) on the degree. 
Since the gauged linear $\s$ models based on these weights
have a Landau--Ginzburg phase we could use Vafa's formulas to compute all
charge degeneracies of Ramond ground states. For the 
109308 reflexive  weights in this class we could thus check the 
coincidence of these numbers with Batyrev's result for the cohomology of
the Calabi--Yau hypersurfaces in the toric varieties defined by the 
reflexive polytopes.
The subtle connections between the two types of Hodge numbers are 
discussed in \cite{ba94}.
Going from the weighted projective space to the variety defined by
a maximal triangulation of 
the reflexive polyhedron we do not resolve all of the singularities
of the ambient space. 
As we saw in our example at the end of section 2, even in cases
when there exists an ``obvious'' way of blowing up the embedding 
variety, this need not be the one leading to the same Hodge numbers.

After the general statistics of transversal and reflexive weights is tables 1 
and 2 we provide lists of weights that may be of particular interest because 
of small Hodge numbers $h_{11}$ or $h_{13}$, or because of a negative Euler
number, which is desirable in the context of SUSY breaking \cite{break}.
In table 3 we give the possible negative Euler number that arise in our lists.
The value $\c=-30$ only occurs for non-reflexive weights, and our smallest 
value $\c=-252$ (which is not divisible by 24) occurs at degree 108 in the 
reflexive case and only at degree 484 in the transversal case.
In table 4 we give the numbers of weights of various types that we find with 
$h_{11}\le7$ or $h_{13}\le7$.
In table 5 we give all 30 transversal and 113 reflexive weights with $\c<0$
and $d\le150$; there is an overlap of 26 weights in this table that are both 
reflexive and transversal. Up to degree 4000 we found 174 more transversal 
weights with negative Euler number, so that altogether we have 291 weights
with $\c<0$
(the smallest values of $h_{11}$ and $h_{13}$ in this list are both 22).

In tables 6 and 7 we list all our weights with  
$h_{11}=1$ or $h_{13}=1$; Fermat weights all have $\c>0$ and
do not contribute to any of our tables of special weights, except for 
8 weights with $6\le d\le 18$ that give $h_{11}=1$.
Some ``first occurrences'' are: \\
The first non-reflexive transversal weight system is $(1,1,1,1,1,2)$.\\
The first non-transversal reflexive weight system is $(1,1,1,1,1,3)$.\\
The first degree that does not admit any weight that is transversal and 
reflexive is 11.\\
The first degree for which no transversal weight system exists is 1733 
(there are only 29 such degrees $d$ in the range $d\le 4000$).

When looking for models with very specific features it may be useful
to go beyond the class of models considered here.
In particular if we are interested in elliptic fibrations it will 
be more ecomonic to generate reflexive polytopes in terms of combined 
weight systems \cite{crp} since the
complete set of relevant weights is already known \cite{wtc}
and the fibrations structure is encoded (and can be pre-selected) in a rather 
simple and explicit way.
Note also that the 3 million elliptic fibrations that were mentioned at
the end of the last section are known to be all connected in a web with
singular transitions that respect the fibration structure because the same is
true for the polytopes of which we are taking direct products \cite{av97}.

\medskip

{\it Acknowledgements.} 
This work is supported in part by the {\it Austrian 
Research Funds} FWF (Schr"odinger fellowship J012328-PHY) and
by the Austrian National Bank under grant Nr. 5674.
We would like to thank P. Candelas, C. Caris, J. Louis, T. Mohaupt, 
E. Perevalov and R. Schimmrigk for discussions.

\newpage

\noindent{\Large\bf Tables}

\def\AF#1&#2&#3{#1&#2&$-$&#3}	\def\ER{\\\hline} \let\ET=\ER \let\ES=\ER

\def\Rlin#1#2#3#4#5{#1&#2&#3&#4&$#5$}			
\def\Tlin#1#2#3#4#5{#1&#2&\expandafter\AF#3&#4&$#5$}
\def\Slin#1#2#3#4#5{{\bf#1}&#2&#3&#4&$#5$}

\bigskip
\begin{center}
\noindent{\bf Table 1:} There are 109308 reflexive weights among the 	\\[2pt]
		\HS35	525572 transversal weights with $d\le4000$.	\\[4mm]
\smallskip\HS-25
\begin{tabular}{||\TVR{4.5}2 c|c|c||c|c|c||c|c|c||c|c|c||} \hline\hline
~d & trans & ref. &d & trans & ref. &d & trans & ref. &d & trans & ref.\\\hline
~100 &	11798 &	3578 &		1100 &	18333 &	3917 &
2100 &	11328 & 2292 &		3100 &	6765  &	1208	\\\hline
~200 &	28457 &	6685 &		1200 &	16351 &	3261 &
2200 &	9694  & 1773 &		3200 &	7449  &	1369	\\\hline
~300 &	31075 &	6716 &		1300 &	14427 &	3045 &
2300 &	8944  & 1627 &		3300 &	6811  &	1300	\\\hline
~400 &	29229 &	6163 &		1400 &	15334 &	3196 &
2400 &	10807 & 2314 &		3400 &	6476  &	1282	\\\hline
~500 &	26792 &	5798 &		1500 &	12907 &	2450 &
2500 &	7385 & 1352 &		3500 &	6209  & 1230	\\\hline
~600 &	26578 &	5649 &		1600 &	13570 & 2764 &
2600 &	9190  & 1897 &		3600 &	6597  &	1329	\\\hline
~700 &	22367 &	4665 &		1700 &	12432 & 2454 &
2700 &	8470  & 1700 &		3700 &	6061  &	1218	\\\hline
~800 &	22139 &	4725 &		1800 &	11594 & 2400 &
2800 &	8134  & 1618 &		3800 &	6288  & 1249	\\\hline
~900 &	20704 &	4478 &		1900 &	11030 & 2118 &
2900 &	7975  & 1523 &		3900 &	5394  &	1046	\\\hline
1000 &	17475 &	3605 &		2000 &	10273 & 1961 &
3000 &	6827  & 1248 &		4000 &	5903  &	1105	\\\hline
\hline
     & 236614 & 52062 &	& 136251 & 27566 &	
     &  88754 & 17344 &	& 63953  & 12336 	\\\hline
\hline\end{tabular}\HS-29 ~
\end{center}

\noindent\bigskip

\begin{center}
        {\bf Table 2:} There are 914051 reflexive, 6918 reflexive and 
	transversal, 	 						\\[2pt]
\HS26	and 25435 general transversal weights with $d\le150$. 		\\[4mm]
\smallskip
\begin{tabular}{||\TVR{4.5}2 c|c|c|c ||} \hline\hline
d  & R		& RT 	& T 	\\\hline
\hline
10 & 9	 	& 8	& 11	\\\hline	
20 & 164 	& 63	& 109	\\\hline	
30 & 835 	& 209	& 422	\\\hline
40 & 2485	& 252	& 684	\\\hline
50 & 5724	& 356	& 1093	\\\hline	
\hline$\sum$ & 9217 & 	888	& 2319	\\\hline
\hline\end{tabular}\HS 19
\begin{tabular}{||\TVR{4.5}2 c|c|c|c ||} \hline\hline
d  & R		& RT 	& T 	\\\hline
\hline
60 &	11489	& 490	& 1480	\\\hline	
70 &	19046	& 375	& 1397	\\\hline	
80 &	30747	& 586	& 2030	\\\hline
90 &	45815	& 744	& 2529	\\\hline
100 &	62779	& 495	& 2043	\\\hline
\hline$\sum$ &	169876  & 2690	& 9479	\\\hline
\hline\end{tabular}\HS 19
\begin{tabular}{||\TVR{4.5}2 c|c|c|c ||} \hline\hline
d  & R		& RT 	& T 	\\\hline
\hline
110 &	84095	& 509	& 2235	\\\hline
120 &	114038	& 955	& 3482	\\\hline
130 &	137806	& 456	& 2005	\\\hline
140 &	178688	& 656	& 2845	\\\hline
150 &	220331	& 764	& 3070	\\\hline
\hline$\sum$ &  734958	& 3340	& 13637	\\\hline
\hline\end{tabular}
\end{center}

\bigskip
\begin{center}
\noindent{\bf Table 3:} Negative values of the Euler number (the value \\[2pt]
\HS49	$\chi=-30$ only occurs for non-reflexive weights).		\\[4mm]
\smallskip\HS-9
\begin{tabular}{||r|r|r|r||r|r|r|r||r|r|r|r||} 
\hline\hline
$-$6 & $-$12 & $-$18	& $-$24 & $-$30 & $-$36 & $-$42 & $-$48  & & $-$60 
						& $-$66 & $-$72\\\hline
   & $-$84 & $-$90  & $-$96 &     &     &     	& $-$120 & & $-$132
						& $-$138 & $-$144\\\hline
   & 	& 	& $-$168 &    & $-$180&     & $-$192 & $-$198 & & &   \\\hline
 	& &	& $-$240 &    & $-$252	&&&&			&&   \\\hline
\hline\end{tabular}\HS-9
\end{center}

\noindent
\begin{center}
        {\bf Table 4:} Numbers of weights with small $1\le h_{11}\le5$ or 
	$1\le h_{13}\le5$ in the list of all reflexive 		\\[2pt]
	weights with $d\le150$, and reflexive (RT) or general (T) transversal 
	weights with  $d\le4000$.					\\[4mm]
\smallskip
\begin{tabular}{||\TVR{4.5}2 c||c|c|c||} \hline\hline
$h_{11}$ & 	R$_{150}$ & RT$_{4000}$ & T$_{4000}$ \\\hline
\hline
1	& 8	& 8	& 33	\\\hline
2	& 33	& 27	& 106	\\\hline
3	& 101	& 66	& 255	\\\hline
4	& 168	& 88	& 411	\\\hline
5	& 267	& 111	& 508	\\\hline
6	& 501	& 183	& 800	\\\hline
7	& 617	& 158	& 789	\\\hline
\hline\end{tabular}
\HS49
\begin{tabular}{||\TVR{4.5}2 c||c|c|c||} \hline\hline
$h_{13}$ &	R$_{150}$ & RT$_{4000}$ & T$_{4000}$ \\\hline
\hline
 1	&$-$	& 6	& 33	\\\hline
 2	&$-$	& 24	& 132	\\\hline
 3	&$-$	& 44	& 196	\\\hline
 4	&$-$	& 48	& 304	\\\hline
 5	&$-$	& 66	& 354	\\\hline
 6	&$-$	& 133	& 533	\\\hline
 7	&5	& 95	& 486	\\\hline
\hline\end{tabular}
\end{center}

\bigskip

\def\betaeta
{	\hline\end{tabular}\HS-9	\end{center}

	\begin{center}
	\smallskip\HS-9		\scriptsize	
\begin{tabular}{||\TVR{4.5}2 c|cccccc||cc|cc|rrr|r|r|r||} \hline\hline
$d$ & $w_1$ & $w_2$ & $w_3$ & $w_4$ & $w_5$ & $w_6$ & $P$ & $V$ & $\5P$ & $\5V$
	& $h_{11}$ & $h_{12}$ & $h_{13}$ & $h_{22}$ & $\chi$	\\\hline
\hline
}

\begin{center}

\noindent~\HS-19
	{\bf Table 5:} Weights for 4-folds with negative Euler 
	number and $d\le150$: Degrees of reflexive {\it and}\HS-19 ~ \\
\HS 9	 transversal weights are in boldface, $V$ and $\5V$ denote the numbers
	 of vertices of $\D$ and $\D^*$, \HS-19 ~  \\
\HS 1	and non-reflexive weights have no entry 
	for the number $\5P$ of lattice points of $\D^*$.\HS 19 ~


\smallskip\HS-9		\scriptsize

\begin{tabular}{||\TVR{4.5}2 c|cccccc||cc|cc|rrr|r|r|r||} \hline\hline
$d$ & $w_1$ & $w_2$ & $w_3$ & $w_4$ & $w_5$ & $w_6$ & $P$ & $V$ & $\5P$ & $\5V$
	& $h_{11}$ & $h_{12}$ & $h_{13}$ & $h_{22}$ & $\chi$	\\\hline
\hline

\Rlin{70} {7&7&7&13&17&19} {84&8&38&7} {26&108&72&220} {-12}\ER
\Rlin{77} {7&7&7&17&19&20} {93&8&38&7} {26&135&81&202} {-120}\ER
\Rlin{80} {5&5&5&14&22&29} {199&8&42&7} {24&210&177&428} {-6}\ER
\Rlin{80} {5&5&5&16&18&31} {199&8&42&7} {24&210&177&428} {-6}\ER
\Rlin{80} {10&13&13&13&15&16} {38&6&48&6} {38&96&26&108} {-144}\ER
\Slin{84} {7&7&7&12&18&33} {133&6&38&6} {22&165&111&246} {-144}\ES
\Rlin{84} {11&11&11&12&18&21} {43&6&38&6} {28&91&31&98} {-144}\ER
\Slin{85} {5&5&5&16&23&31} {214&8&42&7} {24&240&192&428} {-96}\ES
\Rlin{85} {5&5&5&17&21&32} {214&8&42&7} {24&240&192&428} {-96}\ER
\Rlin{88} {8&8&8&19&22&23} {93&8&38&7} {26&135&81&202} {-120}\ER
\Rlin{88} {8&11&12&19&19&19} {44&9&70&8} {63&108&29&196} {-48}\ER
\Slin{90} {5&5&5&18&24&33} {231&6&42&6} {22&272&209&424} {-198}\ES
\Rlin{90} {7&7&7&19&20&30} {84&8&38&7} {26&108&72&220} {-12}\ER
\Slin{90} {9&9&9&10&16&37} {113&6&39&6} {22&144&91&208} {-138}\ES
\Rlin{90} {9&9&9&13&20&30} {84&8&38&7} {26&108&72&220} {-12}\ER
\Rlin{90} {9&9&9&14&19&30} {84&8&38&7} {26&108&72&220} {-12}\ER
\Rlin{90} {9&9&9&17&22&24} {84&8&38&7} {26&108&72&220} {-12}\ER

\betaeta

\Rlin{90} {9&10&14&19&19&19} {48&6&75&6} {65&108&33&220} {-12}\ER
\Rlin{90} {9&12&13&13&13&30} {55&7&38&7} {27&92&42&136} {-90}\ER
\Rlin{90} {11&11&11&15&18&24} {42&6&42&6} {30&101&30&82} {-198}\ER
\Slin{91} {11&13&13&13&16&25} {55&9&34&8} {23&90&43&128} {-96}\ES
\Rlin{92} {7&7&7&22&23&26} {84&8&38&7} {26&108&72&220} {-12}\ER
\Rlin{95} {10&13&15&19&19&19} {42&10&37&9} {30&72&30&140} {-24}\ER
\Rlin{96} {7&7&7&19&24&32} {93&8&38&7} {26&135&81&202} {-120}\ER
\Rlin{96} {7&12&19&19&19&20} {42&10&84&9} {75&126&27&200} {-96}\ER
\Rlin{96} {8&11&19&19&19&20} {42&9&76&8} {69&108&27&212} {-24}\ER
\Rlin{98} {7&7&14&20&24&26} {79&8&39&7} {27&108&69&212} {-24}\ER
\Rlin{99} {7&7&7&22&23&33} {93&8&38&7} {26&135&81&202} {-120}\ER
\Rlin{99} {9&9&9&17&22&33} {93&8&38&7} {26&135&81&202} {-120}\ER
\Rlin{99} {9&9&9&19&20&33} {93&8&38&7} {26&135&81&202} {-120}\ER
\Slin{99} {9&9&9&19&23&30} {93&8&38&7} {26&135&81&202} {-120}\ES
\Rlin{99} {9&10&19&19&19&23} {41&9&91&8} {81&135&26&202} {-120}\ER
\Slin{99} {9&11&11&11&15&42} {108&6&42&6} {24&141&85&198} {-144}\ES
\Rlin{99} {9&11&19&20&20&20} {41&9&91&8} {81&135&26&202} {-120}\ER
\Rlin{99} {9&14&14&14&15&33} {54&6&42&6} {29&102&41&120} {-144}\ER
\Rlin{100} {7&7&7&23&25&31} {93&8&38&7} {26&135&81&202} {-120}\ER
\Slin{102} {6&6&6&17&32&35} {214&8&42&7} {24&240&192&428} {-96}\ES
\Slin{102} {9&12&17&17&17&30} {56&8&34&8} {25&82&43&152} {-36}\ES
\Slin{102} {12&17&17&17&18&21} {44&7&34&7} {26&81&32&114} {-90}\ES
\Rlin{104} {8&13&20&21&21&21} {39&6&93&6} {83&147&24&178} {-192}\ER
\Rlin{105} {7&7&14&24&26&27} {85&9&42&10} {30&126&74&208} {-84}\ER
\Rlin{105} {7&15&17&22&22&22} {45&12&84&9} {74&126&30&208} {-84}\ER
\Rlin{105} {10&15&17&21&21&21} {42&9&37&8} {30&72&30&140} {-24}\ER
\Slin{105} {11&15&15&15&21&28} {55&9&34&8} {23&90&43&128} {-96}\ES
\Rlin{108} {7&7&7&24&27&36} {104&6&38&6} {23&165&92&174} {-252}\ER
\Rlin{110} {5&5&10&22&26&42} {184&8&43&7} {25&200&165&404} {-12}\ER
\Rlin{110} {9&15&20&22&22&22} {44&9&47&8} {36&90&32&136} {-84}\ER
\Rlin{110} {10&11&21&21&21&26} {41&9&91&8} {81&135&26&202} {-120}\ER
\Rlin{110} {10&15&19&22&22&22} {42&9&37&8} {30&72&30&140} {-24}\ER
\Slin{110} {11&11&16&20&22&30} {52&9&34&8} {24&80&42&148} {-36}\ES
\Slin{112} {7&7&14&16&24&44} {119&6&39&6} {23&147&100&242} {-96}\ES

\betaeta

\Rlin{112} {7&16&20&23&23&23} {42&9&92&8} {81&144&27&188} {-168}\ER
\Rlin{112} {8&13&21&21&21&28} {42&9&76&8} {69&108&27&212} {-24}\ER
\Rlin{112} {11&11&16&22&24&28} {39&6&39&6} {29&82&29&112} {-96}\ER
\Rlin{114} {6&14&19&25&25&25} {53&10&88&9} {77&135&38&234} {-72}\ER
\Slin{114} {9&15&19&19&19&33} {55&7&38&7} {27&92&42&136} {-90}\ES
\Slin{114} {15&18&19&19&19&24} {43&7&38&7} {28&91&31&98} {-144}\ES
\Slin{115} {5&5&10&22&31&42} {194&9&45&9} {27&220&174&408} {-66}\ES
\Rlin{115} {5&5&10&23&29&43} {194&9&45&9} {27&220&174&408} {-66}\ER
\Rlin{115} {10&15&21&23&23&23} {42&9&37&8} {30&72&30&140} {-24}\ER
\Slin{120} {5&5&10&24&32&44} {206&6&43&6} {23&242&187&400} {-144}\ES
\Rlin{120} {7&7&14&22&30&40} {79&8&39&7} {27&108&69&212} {-24}\ER
\Rlin{120} {8&8&16&27&30&31} {85&9&42&10} {30&126&74&208} {-84}\ER
\Slin{120} {8&14&15&15&15&53} {111&6&49&6} {29&147&86&210} {-144}\ES
\Rlin{120} {8&15&23&23&23&28} {39&6&93&6} {83&147&24&178} {-192}\ER
\Rlin{120} {10&17&17&18&24&34} {39&7&57&7} {47&92&29&164} {-48}\ER
\Rlin{120} {11&11&20&22&24&32} {38&6&43&6} {31&91&28&98} {-144}\ER
\Rlin{120} {12&13&13&16&26&40} {50&7&39&7} {28&83&39&146} {-48}\ER
\Rlin{121} {10&11&21&21&21&37} {41&9&91&8} {81&135&26&202} {-120}\ER
\Rlin{125} {4&13&25&25&25&33} {87&10&72&8} {56&144&68&252} {-72}\ER
\Rlin{126} {9&9&18&20&28&42} {79&8&39&7} {27&108&69&212} {-24}\ER
\Rlin{126} {9&9&18&22&26&42} {79&8&39&7} {27&108&69&212} {-24}\ER
\Rlin{126} {11&11&11&14&16&63} {113&6&39&6} {22&144&91&208} {-138}\ER
\Rlin{126} {13&13&18&21&26&35} {34&8&55&9} {43&90&23&128} {-96}\ER
\Rlin{128} {7&7&14&30&32&38} {79&8&39&7} {27&108&69&212} {-24}\ER
\Rlin{130} {7&7&7&18&26&65} {199&8&42&7} {24&210&177&428} {-6}\ER
\Tlin{130} {10&10&13&20&24&53} {100&9&9} {26&120&80&228} {-36}\ET
\Rlin{132} {7&7&14&27&33&44} {85&9&42&10} {30&126&74&208} {-84}\ER
\Rlin{132} {8&8&8&31&33&44} {93&8&38&7} {26&135&81&202} {-120}\ER
\Slin{132} {11&11&12&20&22&56} {97&6&43&6} {25&126&77&200} {-96}\ES
\Rlin{132} {11&12&23&23&23&40} {38&6&104&6} {92&165&23&174} {-252}\ER
\Rlin{132} {12&12&12&19&33&44} {93&8&38&7} {26&135&81&202} {-120}\ER
\Slin{132} {12&12&12&23&33&40} {93&8&38&7} {26&135&81&202} {-120}\ES
\Rlin{132} {12&13&13&13&15&66} {108&6&42&6} {24&141&85&198} {-144}\ER
\Slin{132} {12&15&22&22&22&39} {54&6&42&6} {29&102&41&120} {-144}\ES

\betaeta

\Rlin{132} {12&17&17&22&30&34} {34&6&62&6} {52&100&24&148} {-96}\ER
\Rlin{133} {7&7&21&31&33&34} {90&8&49&7} {37&135&78&234} {-72}\ER
\Rlin{134} {7&7&7&22&24&67} {199&8&42&7} {24&210&177&428} {-6}\ER
\Rlin{135} {4&15&27&27&27&35} {87&10&72&8} {56&144&68&252} {-72}\ER
\Rlin{135} {7&7&14&30&32&45} {85&9&42&10} {30&126&74&208} {-84}\ER
\Slin{135} {9&9&18&26&31&42} {85&9&42&10} {30&126&74&208} {-84}\ES
\Rlin{135} {15&19&20&27&27&27} {38&6&48&6} {38&96&26&108} {-144}\ER
\Rlin{136} {7&7&14&31&34&43} {85&9&42&10} {30&126&74&208} {-84}\ER
\Rlin{136} {8&8&8&21&23&68} {214&8&42&7} {24&240&192&428} {-96}\ER
\Slin{136} {16&17&17&24&28&34} {40&7&35&7} {27&73&30&126} {-48}\ES
\Slin{138} {6&6&12&23&44&47} {194&9&45&9} {27&220&174&408} {-66}\ES
\Rlin{138} {6&22&23&29&29&29} {49&6&104&6} {91&165&34&214} {-192}\ER
\Rlin{140} {5&13&28&28&28&38} {83&6&83&6} {64&168&64&220} {-192}\ER
\Rlin{140} {7&7&20&21&30&55} {123&7&49&7} {32&150&102&280} {-48}\ER
\Rlin{140} {10&19&19&26&28&38} {38&6&64&6} {52&104&28&156} {-96}\ER
\Rlin{140} {11&11&20&30&33&35} {40&7&49&7} {39&83&28&146} {-48}\ER
\Tlin{140} {20&20&20&21&24&35} {51&8&7} {27&90&39&128} {-96}\ET
\Rlin{142} {7&7&7&23&27&71} {214&8&42&7} {24&240&192&428} {-96}\ER
\Tlin{143} {10&13&13&19&26&62} {98&10&10} {31&120&76&232} {-30}\ET
\Rlin{144} {7&7&14&32&36&48} {93&6&39&6} {24&147&83&178} {-192}\ER
\Slin{144} {9&9&18&28&32&48} {93&6&39&6} {24&147&83&178} {-192}\ES
\Rlin{144} {9&16&16&21&34&48} {57&6&63&6} {48&114&42&176} {-96}\ER
\Tlin{144} {16&16&18&21&32&41} {48&8&9} {27&84&37&132} {-72}\ET
\Rlin{144} {16&17&17&18&34&42} {33&6&69&6} {57&112&23&140} {-144}\ER
\Rlin{145} {5&5&15&28&39&53} {204&8&54&7} {35&232&183&452} {-36}\ER
\Rlin{145} {5&5&15&29&37&54} {204&8&54&7} {35&232&183&452} {-36}\ER
\Rlin{145} {5&14&29&29&29&39} {83&6&83&6} {64&168&64&220} {-192}\ER
\Rlin{150} {7&7&7&24&30&75} {231&6&42&6} {22&272&209&424} {-198}\ER
\Rlin{150} {10&14&17&17&17&75} {111&6&49&6} {29&147&86&210} {-144}\ER
\Rlin{150} {11&11&25&30&33&40} {39&7&54&7} {42&92&27&136} {-90}\ER
\Rlin{150} {13&13&15&20&39&50} {51&8&49&8} {38&85&38&178} {-6}\ER
\Slin{150} {21&24&25&25&25&30} {42&6&42&6} {30&101&30&82} {-198}\ES
\del
\Slin{152} {12&19&19&20&38&44} {50&7&39&7} {28&83&39&146} {-48}\ES
\Slin{152} {19&19&20&24&32&38} {39&7&39&7} {29&82&29&112} {-96}\ES
\Tlin{154} {11&14&14&20&28&67} {98&10&10} {31&120&76&232} {-30}\ET
\Slin{165} {11&11&15&25&33&70} {100&6&48&6} {30&132&78&212} {-96}\ES
\Tlin{165} {12&17&33&33&33&37} {46&9&8} {36&90&34&144} {-72}\ET
\Slin{168} {7&7&24&28&36&66} {133&6&49&6} {32&166&110&280} {-96}\ES
\Slin{170} {10&10&10&23&32&85} {214&8&42&7} {24&240&192&428} {-96}\ES
\Slin{170} {17&17&20&30&35&51} {41&8&44&8} {36&74&29&156} {-6}\ES

\betaeta

\Slin{175} {6&13&25&25&25&81} {146&9&69&7} {45&180&111&308} {-96}\ES
\Slin{180} {5&5&20&36&48&66} {231&6&54&6} {33&273&208&462} {-144}\ES
\Slin{180} {5&10&15&36&48&66} {156&6&45&6} {25&182&143&352} {-36}\ES
\Slin{180} {9&9&20&32&36&74} {113&6&50&6} {32&146&90&240} {-96}\ES
\Slin{180} {9&18&18&35&40&60} {75&6&48&6} {33&108&65&220} {-12}\ES
\Slin{180} {10&17&36&36&36&45} {48&6&39&6} {30&78&36&152} {-24}\ES
\Slin{180} {12&12&24&31&45&56} {85&9&42&10} {30&126&74&208} {-84}\ES
\Tlin{180} {15&16&36&36&36&41} {46&9&8} {36&90&34&144} {-72}\ET
\Slin{182} {7&12&26&26&26&85} {146&9&69&7} {45&180&111&308} {-96}\ES
\Slin{189} {7&13&27&27&27&88} {143&6&73&6} {47&195&108&274} {-192}\ES
\Slin{190} {10&18&19&19&38&86} {108&6&58&6} {36&144&84&236} {-96}\ES
\Slin{190} {15&19&19&25&55&57} {51&8&49&8} {38&85&38&178} {-6}\ES
\Slin{190} {19&19&25&30&40&57} {40&8&49&8} {39&83&28&146} {-48}\ES
\Tlin{192} {16&16&31&37&44&48} {48&7&7} {34&90&36&144} {-72}\ET
\Slin{198} {15&18&22&22&22&99} {108&6&42&6} {24&141&85&198} {-144}\ES
\Slin{200} {25&25&28&32&40&50} {38&6&43&6} {31&91&28&98} {-144}\ES
\Slin{204} {17&17&24&36&42&68} {44&7&44&7} {35&82&31&144} {-48}\ES
\Tlin{210} {24&25&35&42&42&42} {39&8&7} {39&90&27&128} {-96}\ET
\Slin{216} {16&20&27&27&54&72} {50&7&39&7} {28&83&39&146} {-48}\ES
\Tlin{220} {16&17&44&44&44&55} {46&9&8} {36&90&34&144} {-72}\ET
\Slin{220} {20&22&22&25&65&66} {50&7&54&7} {41&94&37&168} {-48}\ES
\Tlin{225} {22&25&25&29&49&75} {39&8&8} {37&84&27&132} {-72}\ET
\Slin{228} {18&19&19&30&66&76} {55&7&49&7} {37&94&41&168} {-48}\ES
\Slin{228} {19&19&30&36&48&76} {43&7&49&7} {38&92&30&132} {-96}\ES
\Slin{230} {10&10&20&31&44&115} {194&9&45&9} {27&220&174&408} {-66}\ES
\Slin{231} {11&11&21&35&55&98} {117&7&67&7} {46&153&91&286} {-48}\ES
\Slin{240} {5&5&30&48&64&88} {273&6&67&6} {44&322&246&560} {-144}\ES
\Slin{240} {15&15&16&28&60&106} {111&6&63&6} {42&150&84&248} {-96}\ES
\Slin{240} {16&32&39&45&48&60} {37&8&42&9} {30&72&30&140} {-24}\ES
\Tlin{240} {20&20&37&48&55&60} {48&7&7} {34&90&36&144} {-72}\ET
\Slin{246} {6&6&30&41&80&83} {243&8&67&7} {46&280&218&540} {-48}\ES
\Slin{252} {7&7&36&49&54&99} {170&7&73&7} {53&210&141&400} {-48}\ES
\Slin{252} {31&35&36&36&42&72} {34&8&55&9} {43&90&23&128} {-96}\ES
\Slin{255} {7&15&51&51&51&80} {87&10&72&8} {56&144&68&252} {-72}\ES
\Slin{264} {11&11&24&40&66&112} {127&6&67&6} {45&168&99&284} {-96}\ES

\betaeta

\Slin{264} {20&24&33&33&66&88} {49&6&43&6} {30&92&38&132} {-96}\ES
\Tlin{270} {8&15&54&54&54&85} {86&9&7} {59&150&67&248} {-96}\ET
\Slin{270} {20&25&27&27&81&90} {51&8&49&8} {38&85&38&178} {-6}\ES
\Tlin{273} {13&13&33&48&75&91} {69&8&8} {45&114&53&208} {-48}\ET
\Slin{276} {12&22&23&23&69&127} {121&6&74&6} {50&165&91&278} {-96}\ES
\Slin{280} {7&7&40&56&60&110} {183&6&73&6} {52&228&152&404} {-96}\ES
\Tlin{280} {14&19&40&40&80&87} {59&8&9} {53&114&45&208} {-48}\ET
\Tlin{286} {19&20&26&26&52&143} {98&10&10} {31&120&76&232} {-30}\ET
\Slin{288} {9&18&45&56&64&96} {75&7&54&7} {39&117&66&230} {-24}\ES
\Tlin{288} {27&29&32&32&72&96} {39&8&8} {37&84&27&132} {-72}\ET
\Slin{292} {12&21&40&73&73&73} {47&12&70&9} {62&108&32&204} {-36}\ES
\Slin{292} {12&28&33&73&73&73} {44&9&70&8} {63&108&29&196} {-48}\ES
\Slin{294} {20&24&49&49&54&98} {38&9&61&8} {50&96&28&164} {-60}\ES
\Slin{297} {8&17&33&33&66&140} {152&10&94&8} {66&192&114&380} {-24}\ES
\Slin{300} {25&25&42&48&60&100} {42&6&54&6} {41&102&29&120} {-144}\ES
\Slin{300} {25&42&48&50&60&75} {30&6&45&6} {33&71&24&130} {-36}\ES
\Slin{306} {9&16&34&34&68&145} {152&10&94&8} {66&192&114&380} {-24}\ES
\Slin{315} {9&9&61&63&71&102} {128&7&73&7} {58&198&112&328} {-120}\ES
\Slin{315} {9&17&35&35&70&149} {150&8&98&8} {68&204&112&356} {-96}\ES
\Slin{324} {12&17&52&81&81&81} {54&7&75&7} {66&117&39&230} {-24}\ES
\Slin{324} {18&29&34&81&81&81} {48&6&75&6} {65&108&33&220} {-12}\ES
\Slin{330} {22&22&25&30&66&165} {100&6&48&6} {30&132&78&212} {-96}\ES
\Slin{330} {25&30&33&33&99&110} {50&7&54&7} {41&94&37&168} {-48}\ES
\Slin{350} {12&13&50&50&50&175} {146&9&69&7} {45&180&111&308} {-96}\ES
\Slin{360} {9&9&40&64&90&148} {178&6&87&6} {61&228&143&404} {-96}\ES
\Slin{360} {9&9&70&72&80&120} {143&6&73&6} {54&228&126&308} {-240}\ES
\Slin{364} {24&28&39&91&91&91} {42&9&76&8} {69&108&27&212} {-24}\ES
\Slin{366} {14&22&61&61&86&122} {54&7&74&7} {61&120&43&220} {-48}\ES
\Slin{378} {13&14&54&54&54&189} {143&6&73&6} {47&195&108&274} {-192}\ES
\Tlin{384} {17&18&61&96&96&96} {52&10&9} {78&135&37&234} {-72}\ET
\Slin{385} {5&5&55&76&103&141} {385&7&108&7} {78&450&348&848} {-96}\ES
\Slin{388} {21&24&52&97&97&97} {42&9&92&8} {81&144&27&188} {-168}\ES
\Slin{390} {5&5&55&78&104&143} {392&7&108&7} {76&462&355&844} {-138}\ES
\Slin{400} {16&16&65&75&100&128} {73&7&83&7} {68&144&56&252} {-72}\ES
\Slin{400} {30&33&37&100&100&100} {41&9&91&8} {81&135&26&202} {-120}\ES

\betaeta

\Slin{404} {17&20&64&101&101&101} {49&7&96&7} {84&150&34&216} {-144}\ES
\Slin{414} {18&22&69&69&98&138} {53&6&81&6} {66&132&42&212} {-96}\ES
\Slin{420} {5&5&60&84&112&154} {416&6&108&6} {75&492&377&868} {-192}\ES
\Slin{420} {12&12&71&84&105&136} {128&7&73&7} {58&198&112&328} {-120}\ES
\Tlin{420} {19&21&60&60&120&140} {59&8&9} {53&114&45&208} {-48}\ET
\Slin{426} {6&6&60&71&140&143} {361&8&108&7} {79&420&326&824} {-42}\ES
\Slin{435} {14&15&87&87&87&145} {83&6&83&6} {64&168&64&220} {-192}\ES
\Slin{440} {33&37&40&110&110&110} {41&9&91&8} {81&135&26&202} {-120}\ES
\Slin{460} {20&22&73&115&115&115} {49&6&104&6} {91&165&34&214} {-192}\ES
\Slin{462} {6&6&66&77&152&155} {385&7&108&7} {78&450&348&848} {-96}\ES
\Slin{462} {22&22&35&42&110&231} {117&7&67&7} {46&153&91&286} {-48}\ES
\Slin{484} {37&40&44&121&121&121} {38&6&104&6} {92&165&23&174} {-252}\ES
\Slin{500} {13&15&97&125&125&125} {78&11&116&9} {103&180&59&332} {-60}\ES
\Slin{516} {13&16&100&129&129&129} {75&8&121&8} {108&192&56&316} {-120}\ES
\Slin{540} {20&20&81&104&135&180} {77&7&83&7} {67&150&59&248} {-96}\ES
\Slin{544} {12&19&105&136&136&136} {77&10&127&8} {112&198&58&328} {-120}\ES
\Slin{552} {23&23&24&44&184&254} {179&6&125&6} {88&244&132&436} {-96}\ES
\Slin{560} {15&16&109&140&140&140} {77&10&127&8} {112&198&58&328} {-120}\ES
\Slin{564} {13&19&109&141&141&141} {74&7&132&7} {117&210&55&312} {-180}\ES
\Slin{570} {19&19&30&54&190&258} {188&6&128&6} {90&254&140&456} {-96}\ES
\Slin{580} {13&20&112&145&145&145} {74&7&132&7} {117&210&55&312} {-180}\ES
\Slin{594} {16&17&66&66&132&297} {152&10&94&8} {66&192&114&380} {-24}\ES
\Slin{612} {17&17&36&64&204&274} {204&6&136&6} {96&273&153&494} {-96}\ES
\Slin{620} {15&16&124&155&155&155} {78&11&116&9} {103&180&59&332} {-60}\ES
\Slin{624} {16&19&121&156&156&156} {73&6&143&6} {126&228&54&308} {-240}\ES
\Slin{630} {17&18&70&70&140&315} {150&8&98&8} {68&204&112&356} {-96}\ES
\Slin{630} {18&34&35&35&210&298} {198&6&144&6} {102&272&146&492} {-96}\ES
\Tlin{630} {45&45&47&65&113&315} {50&9&9} {76&120&31&232} {-30}\ET
\Tlin{660} {43&54&66&66&101&330} {43&9&9} {80&120&26&228} {-36}\ET
\Slin{680} {15&19&136&170&170&170} {77&10&127&8} {112&198&58&328} {-120}\ES
\Slin{700} {16&19&140&175&175&175} {74&7&132&7} {117&210&55&312} {-180}\ES
\Slin{720} {45&45&48&56&166&360} {63&6&111&6} {84&150&42&248} {-96}\ES
\Slin{732} {28&44&61&61&172&366} {68&7&116&7} {91&153&46&286} {-48}\ES
\Slin{760} {40&45&76&76&143&380} {50&6&113&6} {90&146&32&240} {-96}\ES
\Slin{770} {10&10&103&110&152&385} {385&7&108&7} {78&450&348&848} {-96}\ES

\betaeta

\Tlin{770} {55&55&56&65&154&385} {50&9&9} {76&120&31&232} {-30}\ET
\Slin{780} {19&20&156&195&195&195} {73&6&143&6} {126&228&54&308} {-240}\ES
\Slin{784} {49&49&86&96&112&392} {49&6&111&6} {86&147&29&210} {-144}\ES
\Slin{792} {33&33&72&80&178&396} {74&6&121&6} {91&165&50&278} {-96}\ES
\Slin{808} {34&40&101&101&128&404} {49&7&123&7} {102&150&32&280} {-48}\ES
\Slin{810} {73&80&81&81&90&405} {39&6&113&6} {91&144&22&208} {-138}\ES
\Slin{828} {36&44&69&69&196&414} {67&6&127&6} {99&168&45&284} {-96}\ES
\Slin{855} {14&29&57&57&285&413} {271&8&198&7} {146&364&198&692} {-72}\ES
\Slin{870} {15&28&58&58&290&421} {271&8&198&7} {146&364&198&692} {-72}\ES
\Slin{885} {15&29&59&59&295&428} {269&6&203&6} {149&377&196&670} {-144}\ES
\Slin{900} {25&25&88&116&196&450} {98&8&150&8} {112&204&68&356} {-96}\ES
\Slin{920} {40&44&115&115&146&460} {49&6&133&6} {110&166&32&280} {-96}\ES
\Slin{936} {36&36&115&131&150&468} {70&7&143&7} {111&180&45&308} {-96}\ES
\Slin{952} {28&28&83&132&205&476} {95&9&150&9} {114&192&66&380} {-24}\ES
\Slin{968} {74&80&88&121&121&484} {38&6&133&6} {111&165&22&246} {-144}\ES
\Slin{1053} {13&13&80&139&351&457} {364&7&236&7} {171&474&283&912} {-72}\ES
\Slin{1086} {30&55&96&181&181&543} {49&10&155&9} {140&180&29&360} {-18}\ES
\Slin{1086} {30&66&85&181&181&543} {47&8&155&8} {141&180&27&356} {-24}\ES
\Slin{1092} {13&13&84&144&364&474} {374&6&236&6} {169&492&291&900} {-144}\ES
\Slin{1092} {42&42&131&156&175&546} {70&7&143&7} {111&180&45&308} {-96}\ES
\Slin{1128} {26&38&141&141&218&564} {74&7&169&7} {141&210&53&400} {-48}\ES
\Slin{1134} {14&14&81&160&378&487} {364&7&236&7} {171&474&283&912} {-72}\ES
\Slin{1160} {26&40&145&145&224&580} {74&7&169&7} {141&210&53&400} {-48}\ES
\Slin{1170} {30&38&117&117&283&585} {87&6&178&6} {143&228&61&404} {-96}\ES
\Slin{1260} {36&51&70&70&403&630} {136&6&204&6} {153&273&96&494} {-96}\ES
\Slin{1400} {32&38&175&175&280&700} {74&7&169&7} {141&210&53&400} {-48}\ES
\Slin{1446} {55&60&126&241&241&723} {45&8&205&8} {185&240&25&404} {-132}\ES
\Slin{1518} {55&60&138&253&253&759} {47&10&193&9} {174&220&27&408} {-66}\ES
\Slin{1536} {80&85&91&256&256&768} {44&8&212&8} {192&240&24&428} {-96}\ES
\Slin{1560} {38&40&195&195&312&780} {73&6&183&6} {152&228&52&404} {-96}\ES
\Slin{1566} {54&56&151&261&261&783} {48&6&214&6} {193&252&28&424} {-138}\ES
\Slin{1624} {28&28&114&151&491&812} {199&7&269&7} {198&364&146&692} {-72}\ES
\Slin{1632} {85&91&96&272&272&816} {44&8&212&8} {192&240&24&428} {-96}\ES
\Slin{1680} {48&51&181&280&280&840} {54&6&231&6} {208&273&33&462} {-144}\ES
\Slin{1710} {28&29&114&114&570&855} {271&8&198&7} {146&364&198&692} {-72}\ES

\betaeta

\Tlin{1728} {37&40&211&288&288&864} {69&9&9} {218&280&46&540} {-48}\ET
\Slin{1734} {91&96&102&289&289&867} {42&6&231&6} {209&272&22&424} {-198}\ES
\Slin{1740} {30&30&116&171&523&870} {199&7&269&7} {198&364&146&692} {-72}\ES
\Slin{1740} {56&60&87&87&580&870} {128&6&188&6} {140&254&90&456} {-96}\ES
\Slin{1770} {29&30&118&118&590&885} {269&6&203&6} {149&377&196&670} {-144}\ES
\Slin{1770} {37&42&216&295&295&885} {67&7&251&7} {226&294&44&536} {-96}\ES
\Slin{1974} {42&46&241&329&329&987} {67&6&273&6} {246&322&44&560} {-144}\ES
\Slin{2106} {26&26&139&160&702&1053} {364&7&236&7} {171&474&283&912} {-72}\ES
\Slin{2240} {26&41&160&160&733&1120} {238&8&363&7} {283&474&171&912} {-72}\ES
\Slin{2256} {46&48&141&141&752&1128} {159&6&243&6} {184&322&114&592} {-96}\ES
\Slin{2268} {28&39&162&162&743&1134} {238&8&363&7} {283&474&171&912} {-72}\ES
\Slin{2324} {28&41&166&166&761&1162} {236&6&374&6} {291&492&169&900} {-144}\ES
\Slin{2436} {42&42&151&171&812&1218} {199&7&269&7} {198&364&146&692} {-72}\ES
\Slin{2562} {31&35&361&427&427&1281} {112&10&359&9} {326&420&79&824} {-42}\ES
\Slin{2598} {31&36&366&433&433&1299} {110&8&366&8} {333&432&77&820} {-84}\ES
\Slin{2736} {30&41&385&456&456&1368} {111&9&384&8} {348&450&78&848} {-96}\ES
\Slin{2772} {35&36&391&462&462&1386} {111&9&384&8} {348&450&78&848} {-96}\ES
\Slin{2778} {31&41&391&463&463&1389} {109&7&391&7} {355&462&76&844} {-138}\ES
\Slin{2814} {31&42&396&469&469&1407} {109&7&391&7} {355&462&76&844} {-138}\ES
\Slin{2982} {35&36&426&497&497&1491} {112&10&359&9} {326&420&79&824} {-42}\ES
\Slin{2988} {36&41&421&498&498&1494} {108&6&416&6} {377&492&75&868} {-192}\ES
\Slin{3192} {35&41&456&532&532&1596} {111&9&384&8} {348&450&78&848} {-96}\ES
\Slin{3234} {36&41&462&539&539&1617} {109&7&391&7} {355&462&76&844} {-138}\ES
\Slin{3360} {39&41&240&240&1120&1680} {238&8&363&7} {283&474&171&912} {-72}\ES
\Slin{3486} {41&42&249&249&1162&1743} {236&6&374&6} {291&492&169&900} {-144}\ES
\Slin{3486} {41&42&498&581&581&1743} {108&6&416&6} {377&492&75&868} {-192}\ES
\enddel

\hline\end{tabular}\HS-9

\end{center}

\newpage

\begin{center}

\noindent{\bf Table 6:} Weights for 4-folds with $h_{11}=1$ (for 
	reflexive $\D$ this implies $V=\5V=6$).
\\[4mm]
\smallskip\HS-9		\scriptsize	
\begin{tabular}{||\TVR{4.5}2 c|cccccc||cc|cc|rrr|r|r|r||} \hline\hline
$d$ & $w_1$ & $w_2$ & $w_3$ & $w_4$ & $w_5$ & $w_6$ & $P$ & $V$ & $\5P$ & $\5V$
	& $h_{11}$ & $h_{12}$ & $h_{13}$ & $h_{22}$ & $\chi$	\\\hline

\Slin{6} {1&1&1&1&1&1} {462&6&7&6} {1&0&426&1752} {2610}\ES
\Tlin{7} {1&1&1&1&1&2} {496&10&7} {1&0&455&1868} {2784}\ET
\Slin{8} {1&1&1&1&2&2} {483&6&7&6} {1&0&443&1820} {2712}\ES
\Tlin{9} {1&1&1&1&2&3} {575&10&7} {1&0&523&2140} {3192}\ET
\Slin{10} {1&1&1&1&1&5} {1128&6&8&6} {1&0&976&3952} {5910}\ES
\Tlin{10} {1&1&1&2&2&3} {489&10&7} {1&0&447&1836} {2736}\ET
\Slin{12} {1&1&1&1&2&6} {1167&6&8&6} {1&0&1009&4084} {6108}\ES
\Slin{12} {1&1&1&2&3&4} {603&6&7&6} {1&0&547&2236} {3336}\ES
\Slin{12} {1&1&2&2&3&3} {407&6&7&6} {1&0&373&1540} {2292}\ES
\Slin{14} {1&1&1&2&2&7} {1081&6&8&6} {1&0&935&3788} {5664}\ES
\Tlin{15} {1&1&2&3&3&5} {492&10&7} {1&0&447&1836} {2736}\ET
\Tlin{16} {1&1&1&2&3&8} {1226&9&7} {1&0&1059&4284} {6408}\ET
\Tlin{16} {1&1&2&3&4&5} {509&13&9} {1&0&463&1900} {2832}\ET
\Slin{18} {1&1&2&2&3&9} {984&6&8&6} {1&0&851&3452} {5160}\ES
\Tlin{18} {1&2&3&3&4&5} {309&14&9} {1&0&283&1180} {1752}\ET
\Tlin{20} {1&2&3&4&5&5} {314&10&7} {1&0&287&1196} {1776}\ET
\Tlin{21} {1&1&3&4&5&7} {564&12&9} {1&0&511&2092} {3120}\ET
\Tlin{21} {1&2&3&3&5&7} {378&13&8} {1&0&343&1420} {2112}\ET
\Tlin{22} {1&1&2&3&4&11} {1095&12&8} {1&0&946&3832} {5730}\ET
\Tlin{22} {1&2&3&4&5&7} {358&20&13} {1&0&326&1352} {2010}\ET
\Tlin{24} {2&3&3&4&5&7} {187&13&9} {1&0&171&732} {1080}\ET
\Tlin{26} {1&2&2&3&5&13} {855&12&8} {1&0&739&3004} {4488}\ET
\Tlin{28} {1&1&3&4&5&14} {1148&12&8} {1&0&991&4012} {6000}\ET
\Tlin{28} {1&3&4&5&7&8} {300&17&13} {1&0&273&1140} {1692}\ET
\Tlin{30} {1&2&3&4&5&15} {759&9&7} {1&0&656&2672} {3990}\ET
\Tlin{30} {2&3&4&5&7&9} {189&16&12} {1&0&172&736} {1086}\ET
\Tlin{36} {1&2&3&5&7&18} {896&13&10} {1&0&773&3140} {4692}\ET
\Tlin{40} {1&3&4&5&7&20} {685&14&10} {1&0&591&2412} {3600}\ET
\Tlin{42} {2&3&4&5&7&21} {418&13&9} {1&0&361&1492} {2220}\ET
\Tlin{42} {4&5&6&7&9&11} {93&19&16} {1&0&84&384} {558}\ET
\Tlin{46} {1&4&5&6&7&23} {599&19&13} {1&0&517&2116} {3156}\ET
\Tlin{50} {2&3&5&7&8&25} {419&17&12} {1&0&361&1492} {2220}\ET
\Tlin{60} {3&4&5&7&11&30} {316&14&11} {1&0&271&1132} {1680}\ET
\hline\end{tabular}\HS-9
\end{center}

\newpage

\noindent{\bf Table 7:} Weights for 4-folds with $h_{13}=1$ (in this case
	all polytopes are simplices).

\begin{center}
\smallskip\HS-9		\scriptsize	
\begin{tabular}{||\TVR{4.5}2 c|cccccc||cc|cc|rrr|r|r|r||} \hline\hline
$d$ & $w_1$ & $w_2$ & $w_3$ & $w_4$ & $w_5$ & $w_6$ & $P$ & $V$ & $\5P$ & $\5V$
	& $h_{11}$ & $h_{12}$ & $h_{13}$ & $h_{22}$ & $\chi$	\\\hline

\Tlin{2415} {105&230&279&462&534&805} {7&6&6} {273&0&1&1140} {1692}\ET
\Tlin{2484} {96&265&276&597&621&629} {7&6&6} {273&0&1&1140} {1692}\ET
\Tlin{2520} {180&215&336&364&585&840} {7&6&6} {283&0&1&1180} {1752}\ET
\Tlin{2565} {105&194&410&431&570&855} {7&6&6} {326&0&1&1352} {2010}\ET
\Tlin{2700} {123&240&263&540&675&859} {7&6&6} {273&0&1&1140} {1692}\ET
\Slin{2700} {270&300&369&475&486&800} {7&6&407&6} {373&0&1&1540} {2292}\ES
\Tlin{2777} {335&395&397&407&476&767} {7&6&6} {455&0&1&1868} {2784}\ET
\Tlin{3024} {268&336&384&467&689&880} {7&6&6} {447&0&1&1836} {2736}\ET
\Tlin{3108} {123&279&296&597&777&1036} {7&6&6} {273&0&1&1140} {1692}\ET
\Tlin{3120} {240&260&481&576&715&848} {7&6&6} {447&0&1&1836} {2736}\ET
\Slin{3125} {434&500&520&521&525&625} {7&6&462&6} {426&0&1&1752} {2610}\ES
\Tlin{3216} {171&237&536&609&670&993} {7&6&6} {463&0&1&1900} {2832}\ET
\Tlin{3234} {385&390&474&539&552&894} {7&6&6} {455&0&1&1868} {2784}\ET
\Tlin{3240} {391&407&450&540&558&894} {7&6&6} {455&0&1&1868} {2784}\ET
\Tlin{3240} {396&397&461&474&648&864} {7&6&6} {455&0&1&1868} {2784}\ET
\Tlin{3241} {391&461&463&475&556&895} {7&6&6} {455&0&1&1868} {2784}\ET
\Tlin{3276} {396&403&455&546&576&900} {7&6&6} {455&0&1&1868} {2784}\ET
\Tlin{3360} {240&260&517&672&775&896} {7&6&6} {447&0&1&1836} {2736}\ET
\Tlin{3432} {184&312&429&655&812&1040} {7&6&6} {463&0&1&1900} {2832}\ET
\Tlin{3456} {150&415&432&551&756&1152} {7&6&6} {343&0&1&1420} {2112}\ET
\Slin{3528} {387&432&504&588&637&980} {7&6&483&6} {443&0&1&1820} {2712}\ES
\Slin{3528} {432&441&504&516&631&1004} {7&6&483&6} {443&0&1&1820} {2712}\ES
\Tlin{3582} {350&404&454&597&782&995} {7&6&6} {523&0&1&2140} {3192}\ET
\Tlin{3584} {392&448&456&467&782&1039} {7&6&6} {523&0&1&2140} {3192}\ET
\Tlin{3600} {352&450&464&525&784&1025} {7&6&6} {523&0&1&2140} {3192}\ET
\Tlin{3696} {184&336&439&693&924&1120} {7&6&6} {463&0&1&1900} {2832}\ET
\Slin{3750} {521&600&624&625&630&750} {7&6&462&6} {426&0&1&1752} {2610}\ES
\Tlin{3780} {268&420&480&567&945&1100} {7&6&6} {447&0&1&1836} {2736}\ET
\Tlin{3780} {335&420&480&689&756&1100} {7&6&6} {447&0&1&1836} {2736}\ET
\Tlin{3780} {391&525&540&630&651&1043} {7&6&6} {455&0&1&1868} {2784}\ET
\Tlin{3780} {461&462&540&553&756&1008} {7&6&6} {455&0&1&1868} {2784}\ET
\Tlin{3888} {288&400&436&605&863&1296} {7&6&6} {447&0&1&1836} {2736}\ET
\Slin{3960} {360&396&400&890&891&1023} {7&6&603&6} {547&0&1&2236} {3336}\ES

\hline\end{tabular}\HS-9
\end{center}

\bye